\magnification=1200 \baselineskip=13pt \hsize=16.5 true cm \vsize=23 true cm
\def\parG{\vskip 10pt} \font\bbold=cmbx10 scaled\magstep2

\centerline {\bbold Broad Histogram Monte Carlo}\parG
P.M.C. de Oliveira$^{a}$, T.J.P. Penna$^{a}$ and H.J. Herrmann$^{b}$\par

\item{$a)$} Instituto de F\'\i sica, Universidade Federal Fluminense
av. Litor\^anea s/n, Boa Viagem, Niter\'oi RJ, Brazil 24210-340\par
\item{$b)$} ICA 1, Universit\"at Stuttgart, Pfaffenwaldring 27,
70569 Stuttgart, Germany\par

\vskip 0.5cm
\leftskip=1cm \rightskip=1cm
{\bf Abstract}: We propose a new Monte Carlo technique in which the
degeneracy of energy states is obtained with a Markovian process analogous to
that of Metropolis used currently in canonical simulations. The obtained
histograms are much broader than those of the canonical histogram technique
studied by Ferrenberg and Swendsen. Thus we can reliably reconstruct
thermodynamic functions over a much larger temperature scale also away from
the critical point. We show for the two-dimensional Ising model how our new
method reproduces exact results more accurately and using less computer time
than the conventional histogram method. We also show data in three dimensions
for the Ising ferromagnet and the Edwards Anderson spin glass.\par
\leftskip=0pt \rightskip=0pt\vskip 0.5cm 

In the simulation of thermal systems one typically needs to calculate a
variable $<Q>_T$ as a function of temperature $T$. Usually one would have to
perform independent Monte Carlo simulations at different values of
temperature. An appealing strategy to avoid such multiple simulations is the
canonical histogram method [1,2]: The histogram $P_T(E)$ of the energies at
one given temperature $T_0$ is measured and then the distribution at a
different temperature $T$ is obtained by reweighting, i.e. by multiplying
with $\exp(E/T-E/T_0)$ and normalizing. In order to obtain the average
$<Q>_T$, one needs to accumulate also the histogram of $Q$ as a function of
energy $E$. The thermal average at temperature $T$ is then

$$<Q>_T = \sum_E <Q(E)> {P_{T_0}(E) e^{E(T_0-T)/T_0T}
\over \sum_{E'} P_{T_0}(E') e^{E'(T_0-T)/T_0T}}\,\,\,\,\, ,
\eqno(1)$$

\noindent where $<Q(E)>$ means the average value of $Q$ obtained at fixed
energy $E$. For simplicity we set $k_B = 1$ in this article.

The histogram method has been carefully checked on various models [3]. Its
main disadvantage is that the canonical distribution $P_T(E)$ of the energy
is rather narrowly peaked around the average value $<E>_{T_0}$ (and the more
so the larger the system) so that when $T$ is not close to $T_0$ the
reweighting factor is very small there and very large near $<E>_T$. One also
has strong statistical fluctuations stemming from the tails of the
distributions. So, in order to get reliable results the tails must be sampled
very well by making good statistics and the temperatures considered should
not be far from $T_0$.

We want to introduce here a completely different technique based on the
calculation of the degeneracy $g(E)$ of energy states from histograms of
adequate macroscopic quantities defined below [4]. Figure 1 compares the
classical canonical histogram for an Ising model on square lattices to the
one obtained with our method. As mentioned before the width of the classical
histogram method decreases with system size. Instead the histogram introduced
in the following covers the entire energy range.

The first step of our method [4] is to perform a Markovian walk along the
energy axis which samples all energies with the same weight. For that we
define two classes of moves in phase space:

\hskip 2cm {class 1:$\,\,\,\,\,\,\,\,\,\,  E \,\, $\hbox to 30pt
{\rightarrowfill} $\,\, E - \Delta{E}$}\par
\hskip 2cm {class 2:$\,\,\,\,\,\,\,\,\,\,  E \,\, $\hbox to 30pt
{\rightarrowfill} $\,\, E + \Delta{E}\ \  ,$}\par

\noindent where $\Delta{E} > 0$. To correctly sample $g(E)$ we propose the
following dynamics: We randomly choose a move. If this move belongs to class
1, it is accepted. If, however, it belongs to class 2, it is accepted only
with probability $N_{\rm dn} / N_{\rm up}$, where $N_{\rm dn}$ and $N_{\rm
up}$ are the total numbers of possible moves of classes 1 and 2,
respectively, measured at the current state.

These quantities $N_{\rm dn}$ and $N_{\rm up}$ concern the whole lattice,
considering {\it all possible} movements one can perform at the current state
in order to get a (would-be) next state along the Markovian chain. In the
particular case of a single spin flip dynamics, $N_{\rm dn}$ and $N_{\rm up}$
are obtained by checking on each site if a spin flip would increase or
decrease the energy and counting how many times each case occured. Then one
site is chosen randomly and the above rule is applied only to it.
Nevertheless, the method is not restricted to single spin flips, and any
other updating protocol can be adopted, provided {\it all} possible movements
are classified as belonging to classes 1 or 2, in order to count their
current numbers $N_{\rm dn}$ and $N_{\rm up}$, before actually performing one
of them. Following this rule the probabilities to increase or to decrease the
energy are equal.

Like for a random walk, the region already visited along the energy axis
increases proportionally to $\Delta{E} \sqrt{t}$, where $t$ is the number of
performed moves, i.e. the length of the Markovian sequence of states. This
allows one to obtain any predefined distribution width, simply by taking a
large enough computer time. On the other hand, in canonical dynamics based on
the Boltzmann weights one always get distributions with exponential decaying
tails, and thus with time-independent widths. Figure 1 shows the number of
visits as a function of the energy.

The second step of our method is the calculation of the degeneracy of energy
states $g(E)$. In the above Markovian process, the probability for the energy
to jump from $E$ to $E + \Delta{E}$ is the same as that of jumping back from
$E + \Delta{E}$ to $E$:

$$<N_{\rm up}(E)> g(E) =\,\, <N_{\rm dn}(E+\Delta{E})> g(E+\Delta{E})\,
\eqno(2)$$

\noindent where the averages are taken at fixed energy. Equation (2) can be
rewritten as

$$\beta(E) \equiv {{\rm d} \ln g(E) \over {\rm d}E} = {1 \over \Delta{E}}\,\,
\ln {<N_{\rm up}(E)> \over <N_{\rm dn}(E+\Delta{E})>}\,\,\,\,\, ,
\eqno(3)$$

\noindent In this way one obtains $g(E)$, from the averages $<N_{\rm up}(E)>$
and $<N_{\rm dn}(E)>$ accumulated during the random walk.

Summarizing, the method consists in performing the Markovian process defined
above, and accumulating four histograms along the $E$ axis: for the number of
visits; for the quantity $Q$ one is interested in; for the average number
$N_{\rm dn}$ of moves of class 1; and for $N_{\rm up}$ corresponding to class
2. From these numbers, $g(E)$ is determined by equation (3), and the thermal
average $<Q>_T$ obtained through equation (1) using

$$P_T(E) = {1\over Z_T} g(E) \exp(-E/T)
\eqno(4)$$

\noindent with $Z_T = \sum_E g(E) {\rm exp}(-E/T)$, if one wants to work in
the canonical ensemble.

For the one-dimensional Ising model we have been able to prove that our model
gives the exact solution for $g(E)$ [4]. Also in this same reference one can
find a generalization of equation (3) for the case where different values of
$\Delta E$ are possible: it is enough to transfer the factor $1/\Delta E$ as
an exponent in both the numerator and denominator, inside the average
brackets.

In order to test our new technique numerically we performed simulations for
three examples: the two- and three-dimensional Ising ferromagnet and the 3D
Ising spin glass. Starting with a random configuration we choose in the
ferromagnetic cases an initial state that was thermalized at the critical
temperature for 10 Metropolis lattice sweeps [5]. For the spin glass we did
instead 100 sweeps at zero temperature. We checked, however, that the results
did not depend on the initial state. Our systems had periodic boundary
conditions in all directions. For efficiency we used the multilattice
approach [6] storing 32 lattices in an array of 32-bit integers. All 32
samples are processed in parallel by using multi-spin coding [7].

The degeneracy $g(E)$ of states obtained for the two-dimensional Ising model
on a $32 \times 32$ square lattice agrees so well with the recently derived
exact degeneracy [8] that on a plot they are indistinguishable from each
other. Figure 2 shows the averaged energy and specific heat, obtained by the
present method. The data are extremely close to the exact curves [9].  This
is true even for the specific heat at its maximum where statistical
fluctuations are largest as can be seen from the inset of figure 2.

In figure 2 we also see data obtained using the classical canonical histogram
method [1,2]. The data noticeably deviate around the maximum of the specific
heat which is probably due to the fact that the width of the canonical
histogram is not larger than the distance between this maximum and $T_c$, the
temperature at which the histogram was calculated. Although the computational
effort invested in CPU time was about 50\% larger for our method than for the
canonical histogram method (using in both cases the same multi-spin
implementation) the result obtained with our method is at least ten times
more precise. Moreover, for the canonical histogram technique one must
simulate independently at several temperatures to get the data in the
temperature range considered. This is a convincing evidence for the
efficiency of our method.

For the two-dimensional Ising ferromagnet the magnetization [10] and
susceptibility were also analyzed and compared with canonical Monte Carlo
simulations [11]. The agreement is again excellent. We did the same studies
also for the three-dimensional Ising model. Again energy, specific heat,
magnetization and susceptibility were calculated and found in very good
agreement with known results investing a quite modest computational effort.
In figure 3 we show as an example magnetization and susceptibility for 3D
systems differing in size. For large lattices like the $50 \times 50 \times
50$ presented in figure 3, the random walk dynamics may be very slow, if one
starts always from only one energy. In these cases, it is better to restart
the dynamics many times during the simulation, each time choosing a new value
for the starting energy. This trick was adopted also for the $256 \times 256$
lattice in figure 1.

As a more difficult test we chose also to simulate the three-dimensional
Ising spin glass. Besides energy and specific heat we also calculated the
overlap order parameter defined through

$$q = [{1\over N} \sum_i <s_i s_i^{(0)}>_T]_{av}\quad ,\eqno(6)$$

\noindent where $\{ s_i^{(0)} \}$ is a ground state previously determined by
the approximation introduced in [12]. In figure 4 we show that it goes to
zero close to the predicted temperature [13]. Again the method turns out to
be very efficient.

Two last points deserve comments. First, there are other methods [14] which
can be thought as similar to our first step, where we defined a non-biased
random walk dynamics. All of them introduce non-canonical ensembles directly
weighted by the unknown entropy, or $g(E)$. Karliner et al [14], for
instance, use the $g(E)$ scaled up from exact calculations on small lattices
in order to define the statistical weight of their dynamics, replacing the
Boltzmann factors. In all these earlier works, the entropy or $g(E)$ is
measured directly from the accumulated number of visits to each energy
channel, within the particular dynamics corresponding to each case. Our
approach is distinct from those others: we do not use the number of visits,
on the contrary, our results are extracted from accumulated histograms of
quantities measured within each visited state, namely $N_{\rm dn}$ and
$N_{\rm up}$, concerning the whole lattice. The only role played by the
number of visits to each energy channel is that it must be large enough to
provide a good statistics. Being based on statistical averages of macroscopic
quantities, our method also has a better numerical precision.  Indeed,
accumulating the macroscopic quantities $N_{\rm dn}$ and $N_{\rm up}$ (both
of the order of the number of sites), instead of merely incrementing by unity
a visit counter after each movement as in the traditional histogram method,
we succeed in getting better numerical accuracy.

Second, the left hand side of equation (3) coincides with the formal
definition of (inverse) temperature as a function of the energy $E$. Thus,
the averages on the right hand side of the same equation correspond to the
canonical ensemble equilibrium. Although in our second step we have derived
equation (3) from the particular dynamics defined in the first step (the
non-biased random walk), it holds for any other dynamics respecting the
canonical equilibrium for each (inverse) temperature $\beta(E)$. In this
sense, the importance of our method relies on the second step, the
determination of $g(E)$ from the countings of potential movements $N_{\rm
dn}$ and $N_{\rm up}$, measured within {\it any} canonical equilibrium
dynamics.

We have presented a new histogram Monte Carlo method which as compared to the
traditional one based on temperature [1,2] is based on the histograms of the
energy. These histograms have the advantage of having much broader tails
allowing to extrapolate to a much larger range of temperatures with a rather
small number of samples. We have tested our method on the two- and
three-dimensional Ising model and the 3D Ising spin glass and succeeded in
reproducing thermodynamic quantities with higher accuracy over a larger
temperature scale and with less computational effort than the canonical
histogram method.

We thank L. de Arcangelis, D. Stauffer, S. Moss de Oliveira, C. Moukarzel and
Yi-Cheng Zhang for helpful discussions.

\vskip 30pt
\centerline{\bf References}\parG
\item{[1]} Salzburg Z.W., Jacobson J.D., Fickett W. and Wood W.W., {\it J.
Chem. Phys.} {\bf 30} 65 (1959).\par
\item{[2]} Ferrenberg A.M. and Swendsen R.H., {\it Phys. Rev. Lett.} {\bf
61}, 2635 (1988).\par
\item{[3]} Ferrenberg A.M. and Landau D.P., {\it Phys. Rev.} {\bf B44},
5081 (1991).\par
\item{[4]} P.M.C. de Oliveira, T.J.P. Penna and H.J. Herrmann, {\it Braz. J.
of Physics} {\bf 26}, 677 (1996).\par
\item{[5]} In some cases it was also useful to perform one or several
Metropolis updates after each iteration of the Markov process at the
temperature given by equation (3) in order to reduce correlations and
accelerate the algorithm.\par
\item{[6]} Bhanot G., Duke D. and Salvador R., {\it J. Stat. Phys.} {\bf
44}, 985 (1986); {\it Phys. Rev.} {\bf B33}, 7841 (1986).\par
\item{[7]} de Oliveira P.M.C., {\sl Computing Boolean Statistical Models}
World Scientific, Singapore, ISBN 981-02-0238-5 (1991).\par
\item{[8]} Beale P.D., {\it Phys. Rev. Lett.} {\bf 76}, 78 (1996).\par
\item{[9]} Ferdinand A.E. and Fisher M.E., {\it Phys. Rev.} {\bf 185}, 832
(1969).\par
\item{[10]} In order to break the global spin flip symmetry, the
magnetization was in fact defined as the average of the absolute difference
between the population fractions of spins up and down.\par
\item{[11]} see, for instance, Landau L.D., {\it Phys. Rev.} {\bf B13}, 2997
(1976), or de Oliveira P.M.C. and Penna T.J.P., {\it Rev. Bras. F\'\i s.}
{\bf 18}, 502 (1988).\par
\item{[12]} Stauffer D. and de Oliveira P.M.C. {\it Physica} {\bf A215}, 407
(1995).\par
\item{[13]} A.T. Ogielski, Phys. Rev. B {\bf 32}, 7384 (1985), N. Kawashima
and A.P. Young, Phys. Rev. B {\bf 53}, R484 (1996).\par
\item{[14]} B. Bhanot, S. Black, P. Carter and S. Salvador, {\it Phys. Lett.}
{\bf B183}, 381 (1987); M. Karliner, S. Sharpe and Y. Chang, {\it Nucl. Phys.}
{\bf B302}, 204 (1988); B. Berg and T. Neuhaus, {\it Phys. Lett.} {\bf B267},
249 (1991); {\it Phys. Rev. Lett.} {\bf 68}, 9 (1992); B. Berg, {\it Int. J.
Mod. Phys.} {\bf C3}, 1083 (1992); B. Berg, T. Celik and U. Hansmann, {\it
Europhys. Lett.} {\bf 22}, 63 (1993); J. Lee, {\it Phys.  Rev. Lett.} {\bf
71}, 211 (1993); B. Hesselbo and R.B. Stinchcombe, {\it Phys. Rev. Lett.}
{\bf 74}, 2151 (1995).\par

\vfill\eject
\centerline{\bf Figure Captions}\vskip 50pt

\item{Figure 1} Number of visits as a function of energy obtained from the
traditional histogram method [1,2] (fixing the temperature at the critical
value), and from the present method, for the Ising ferromagnet on $32 \times
32$ and $256 \times 256$ square lattices. The energy is scaled such that the
critical value corresponds to 0.146. Energies above $\sim 0.4$ are not
sampled for the $256 \times 256$ lattice, because they do not contribute to
thermal averages at the interesting range of temperatures ($0 < T <
2T_c$).\parG

\item{Figure 2} Averaged energy and specific heat obtained from the present
method for the Ising ferromagnet on a $32 \times 32$ square lattice and the
exact ones [8] (both full lines). Except for a small region around the
critical point (see inset for maximum of specific heat) the two curves are
indistinguishable. Results of similar quality were also obtained for larger
lattices (we tested $L = 64$, $128$ and $256$). The open circles represent
data obtained using the canonical histogram method [1,2] for which $P_T(E)$
was obtained at $T_c$. Our method needed just 40 minutes on a Pentium (66MHz)
for the $6\times 10^4$ lattice sweeps performed.\parG

\item{Figure 3} Magnetic susceptibility obtained from the present method for
the 3D Ising ferromagnet on a $50 \times 50 \times 50$ (upper curves, showing
also the magnetization), a $20 \times 20 \times 20$ (middle curve) and a $10
\times 10 \times 10$ (lower curve) cubic lattice.\parG

\item{Figure 4} Overlap with one specific previously calculated low
temperature state and non-linear susceptibility of this order parameter for
the 3D Ising spin glass simulated on a $8 \times 8 \times 8$ cubic lattice
making 53000 lattice sweeps for 128 different configurations of couplings.
The total amount of CPU time spent to get these data was 15 hrs on a Pentium
(66MHz).\parG

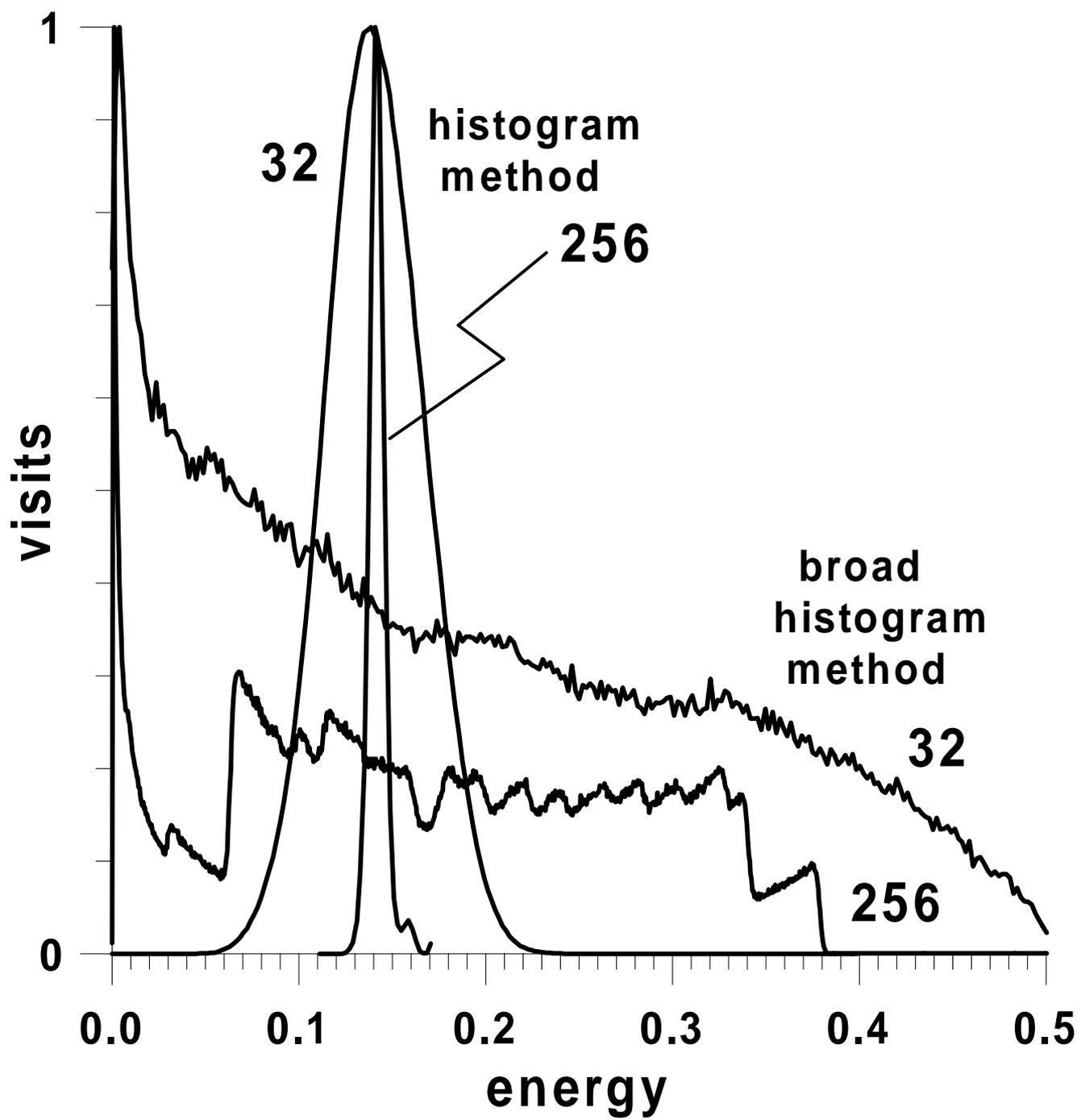
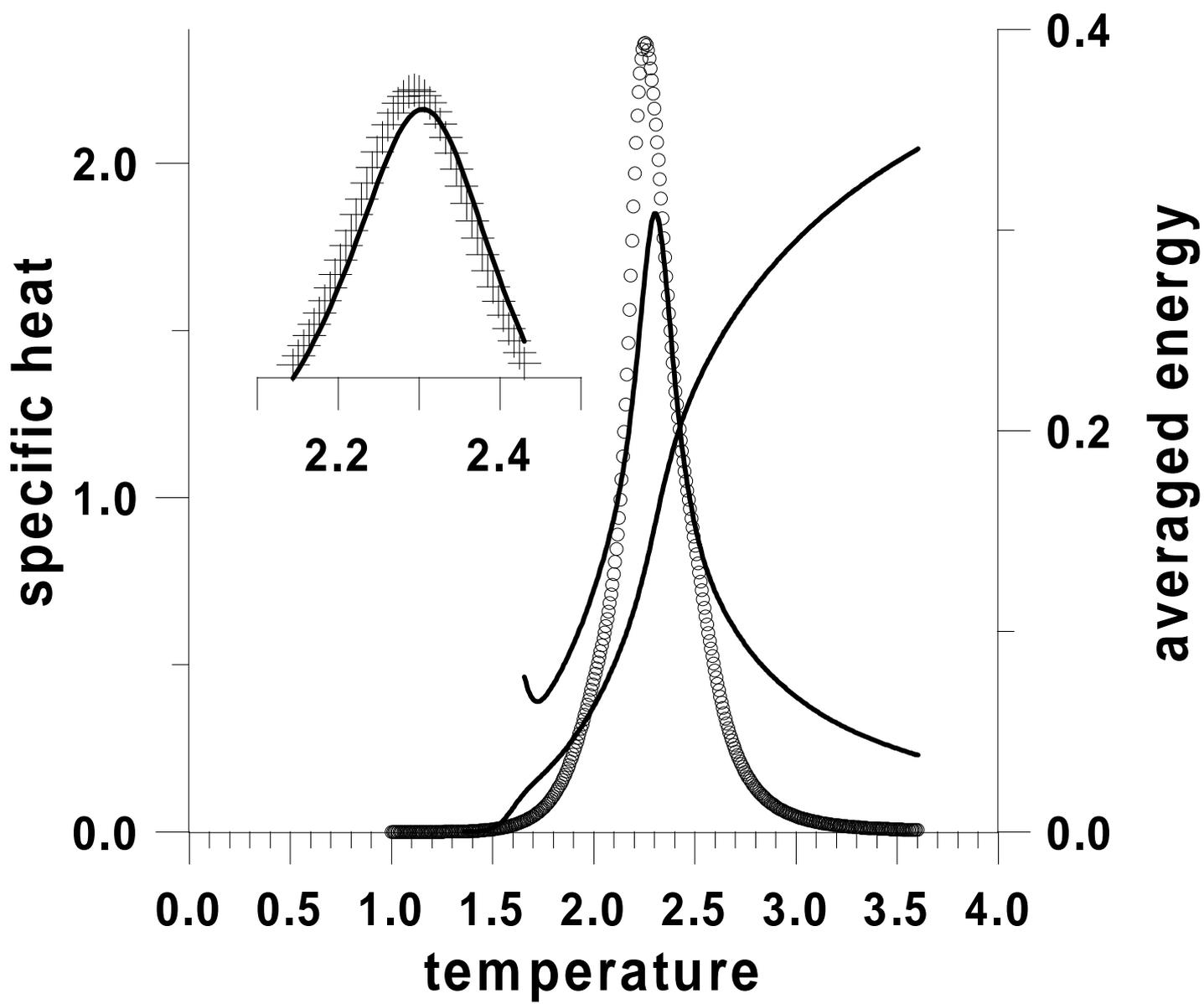
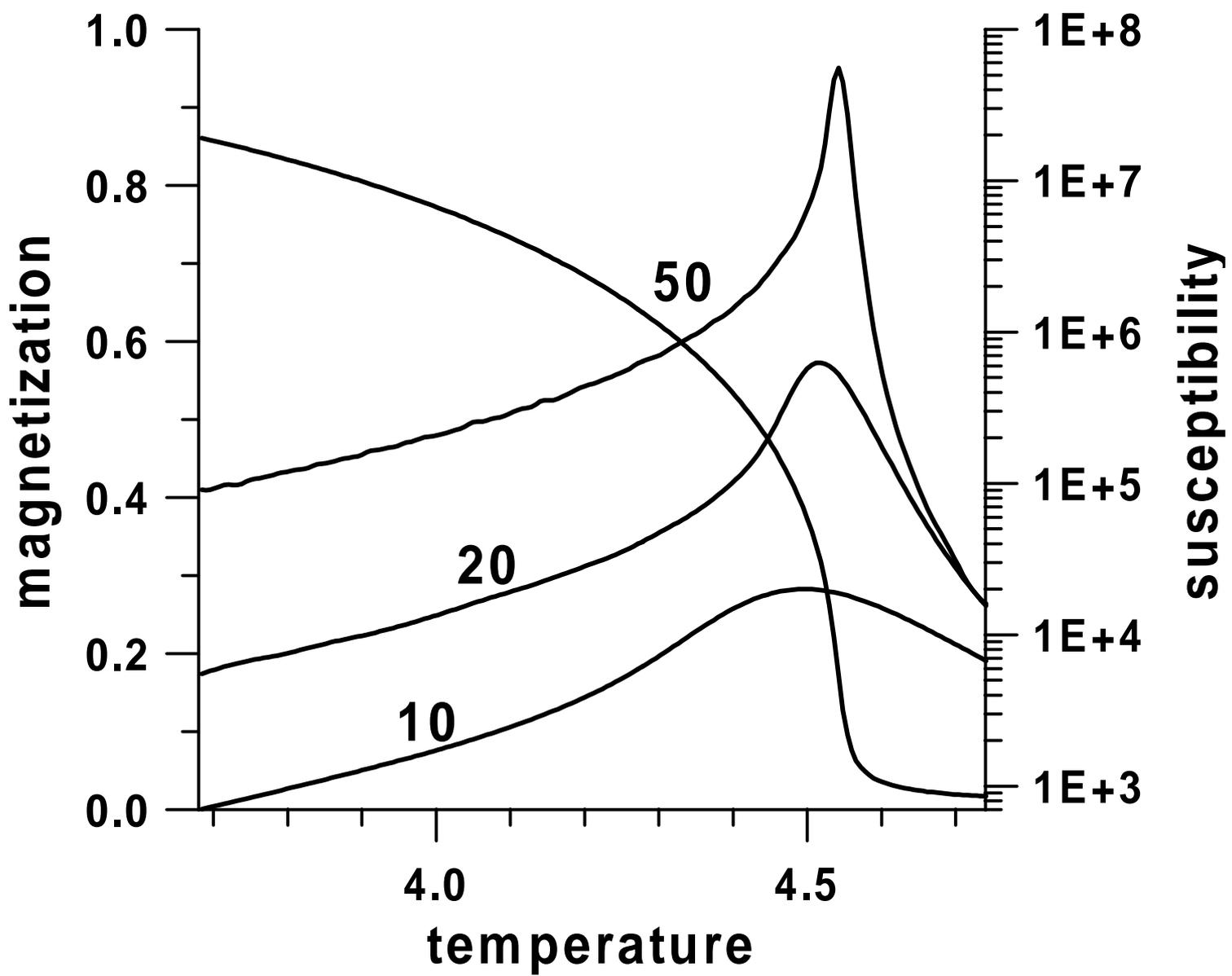
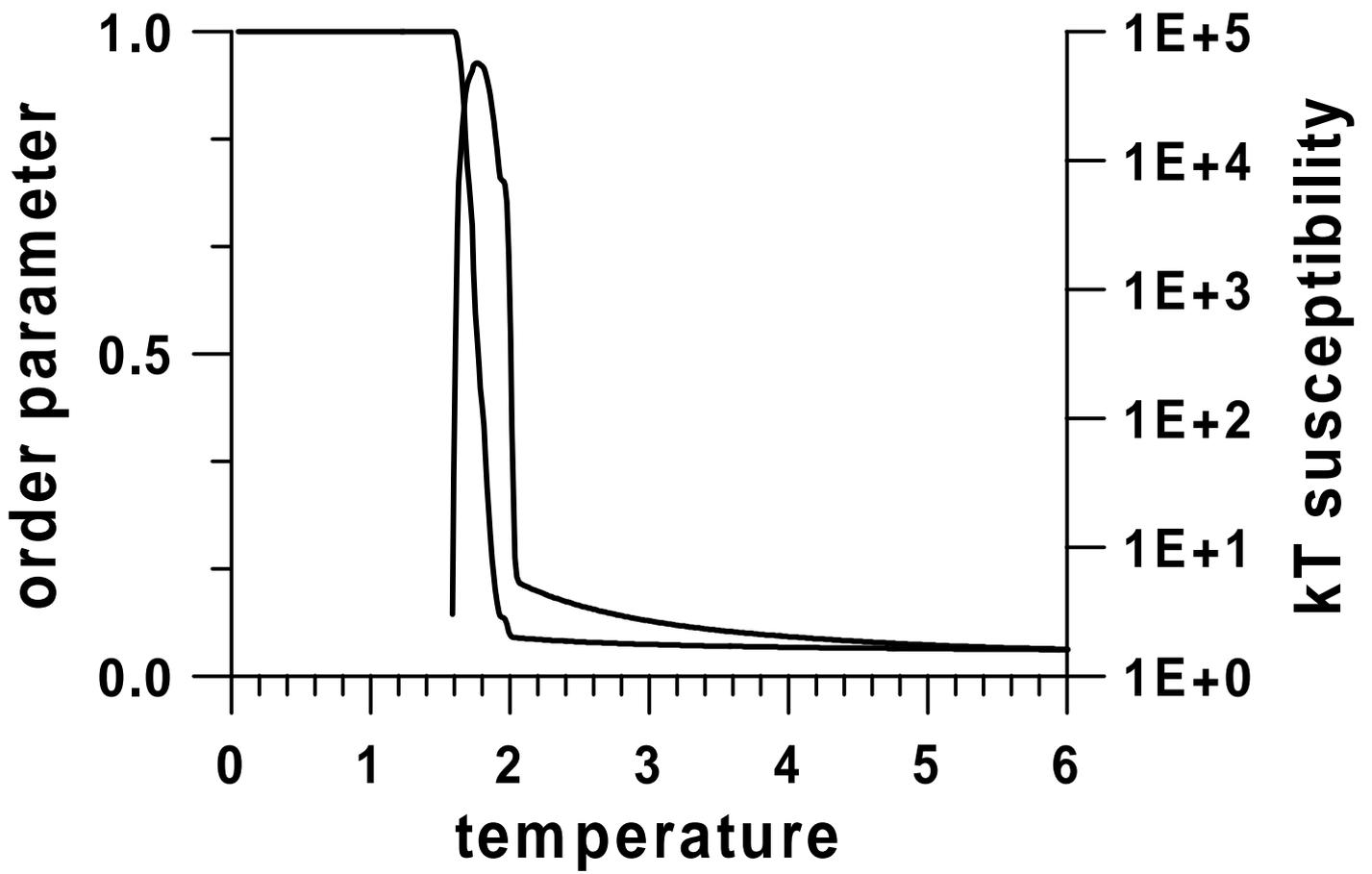
\bye